# Impact of nitrogen incorporation on interface states in (100)Si/HfO$_2$

Y. G. Fedorenko, L. Truong, V. V. Afanas'ev, and A. Stesmans

Department of Physics, University of Leuven, Celestijnenlaan 200 D, 3001 Leuven, Belgium

Z. Zhang and S. A. Campbell

Department of Electrical and Computer Engineering, University of Minnesota, Minneapolis, MN 55455, USA

**Abstract**

The influence of nitrogen incorporation on the energy distribution of interface states in the (100)Si/HfO$_2$ system and their passivation by hydrogen have been studied. The results are compared to those of nominally N-free samples. The nitrogen in the (100)Si/HfO$_2$ entity is found to increase the trap density, most significantly, in the upper part of Si band gap, in which energy range nitrogen incorporation prevents passivation of interface traps by hydrogen. At the same time, passivation of fast interface traps in the lower part of the band gap proceeds efficiently, provided the thickness of the nitrogen containing interlayer is kept within a few monolayers. The minimal interface trap density below the midgap achieved after passivation in H$_2$ is dominated by the presence of slow N-related states, likely located in the insulator. As inferred from capacitance-voltage and ac conductance analysis, the lowest density of electrically active defects [$(8-9) \times 10^{10}$ eV$^{-1}$cm$^{-2}$ at 0.4-0.5 eV from the top of the Si valence band edge] is achieved both in the N-free and N-containing (100)Si/HfO$_2$ structuresafter post-deposition anneal at 800 $^0$C in (N$_2$+5% O$_2$) followed by passivation in molecular hydrogen at 400 $^0$C for 30 min.

## I. Introduction

Hafnium oxides containing nitrogen receive much attention because of several potential benefits the N incorporation may provide: N-containing insulators on Si represent an effective diffusion barrier during subsequent high temperature processing, nitrogen helps to prevent crystallization of the metal oxide, and it also enhances the dielectric permittivity of the interfacial layer.[1-5] Pre-grown silicon nitride ($Si_3N_4$) can also be used to promote two-dimensional nucleation of subsequently deposited $HfO_2$ and, at the same time, inhibit interlayer formation by oxidation of silicon.

Nitrogen can be introduced by employing different technological approaches, like remote plasma processing including monolayer interface nitridation[6], pre-deposition of $Si_3N_4$ or thermal nitridation[3] to form a silicon nitride interlayer, post-deposition anneal in $NH_3$ of sputtered hafnium silicate films[4, 5], admixing of nitrogen monoxide (NO) to the hafnium precursor used in metal organic chemical vapor deposition (MOCVD) technology, or, by using a N-containing precursor such as $Hf(NO_3)_4$ in nitrato-CVD process (NCVD)[1]. As revealed by electron spin resonance (ESR) analysis[7] the deposition of $HfO_2$ using the $Hf(NO_3)_4$ precursor results in some incorporation of N in the metal oxide network, inherent to the particular CVD process. Potentially, the presence of nitrogen in the interfacial region and in the hafnia itself may result in a higher density of interface traps and hinder their passivation.

Naturally, the question arises of how N-associated defects may affect the electrical performance, which is linked to the interface properties, in particular, the interface trap density ($D_{it}$). In the present work we analyse $D_{it}$ in the (100)Si/$HfO_2$ interface with nitrogen incorporated by different technological methods. The effect of postdeposition annealing (PDA) treatments in oxygen ambient and forming gas ($H_2$) on $D_{it}$ is also studied. It is found that the $D_{it}$ in as-grown samples as well as their passivation in molecular hydrogen is sensitive to the amount of nitrogen incorporated in the interfacial layer (IL) and to the method of nitridation. The presence of thin silicon nitride (0.3-0.4 nm) pre-grown in ammonia at 600 °C prior to HfO$_2$ deposition by CVD does not hamper the passivation of fast interface states below silicon midgap. The donor-like traps in those samples can be efficiently passivated by molecular hydrogen as is the case for the N-free (100)Si/HfO$_2$ interface. The lowest $D_{it}$ in the N-containing samples is achieved after passivation in molecular hydrogen. The remaining trap density is dominated by the presence of deep trap levels located at 0.4-0.5 eV from the top of the silicon valence band. For all studied interfaces the most efficient passivation has been universally observed to occur at 400 $^0$C as in the case of (100)Si/SiO$_2$. As compared to the latter, annealing in H$_2$ at 550 $^0$C results generally in a 2 to 3 times higher $D_{it}$ values than that obtained at 400 $^0$C. However, the beneficial aspect of nitrogen incorporation appears to occur at the expense of a higher density of acceptor-like interface traps.

## II. Experimental

Interface states and their passivation by hydrogen were studied in (100)Si/HfO$_2$ structures with nitrogen incorporated by several methods. The reference N-free samples were prepared by atomic-layer CVD (ALCVD) using H$_2$O and HfCl$_4$ precursors on (100) silicon covered with a chemical oxide layer (SiO$_x$) grown in ozonated water. Further, these were compared to several N-containing structures: The substrate of the same type were used to deposit HfO$_2$, using O$_2$ and tetrakis-dimethylaminohafnium metallo-organic (MO)-CVD in which nitrogen is present in a form of amino groups. (1) (100)Si/Si$_3$N$_4$/HfO$_2$ entities with different thicknesses of the Si$_3$N$_4$ interlayer (IL) grown in ammonia (10$^{-5}$ Torr) for 2 min at 600 or 900 °C and with a HfO$_2$ layer on top grown by CVD using hafnium tert-butoxide (HTB) with and without addition of NO; (2) NCVD grown (100)Si/SiON/HfO$_2$ structures in which the oxynitride growth resulted from Hf(NO$_3$)$_4$ decomposition byproducts, i.e., NO and NO$_2$, in nominally H and C free conditions, unlike the above methods; (3) NCVD grown (100)Si/Si$_3$N$_4$/HfO$_2$ structures with a $Si_3N_4$ IL pre-grown in ammonia for 2 min at 900 °C as in case (1). Some parameters of the studied samples are compiled in Table 1. More details are given elsewhere.[8,9]

Metal-oxide-silicon (MOS) capacitors were prepared by thermoresistive evaporation of Au electrodes of $(1.3-1.6) \times 10^{-3}$ $cm^2$ area. No post-metallization anneal was performed to ensure negligible passivation of interface traps by hydrogen. Interface trap densities were determined in as-deposited samples (additionally exposed to 10-eV photons at room temperature to photodissociate hydrogen from passivated Si dangling bonds[10]), in the samples subjected to post-deposition anneal (PDA) in $N_2$+ 5% $O_2$ at 800 °C for 10 min, and in the samples subsequently passivated in $H_2$ at 400 °C or 550 °C for 30 and 10 min, respectively. A fresh sample was used for each of those treatments. The $D_{it}$ values were determined using the ac conductance method following Nicollian and Brews[11] and the Berglund capacitance-voltage (CV) technique[12]. When using the Berglund procedure to reliably determine the $D_{it}$ distributions within the about 80% central part of the band gap, corona charging was applied to the capacitors prior to CV curve recording.[13] Comparison of $D_{it}$ distributions obtained by those two methods allowed us to evaluate the relative contributions of slow and fast interface states to the observed trap density.

**III. Results and discussion**

   **A. As-grown samples**

Figure 1 shows $D_{it}$ data determined by the ac conductance method for as-grown N-free (ALCVD sample D:▽) and N-containing samples. Four methods of nitridation are compared: Pre-growth of a $Si_3N_4$ IL and subsequent CVD deposition of $HfO_2$ using HTB with (1) NO (A1: ○, A2: □) or without (2) it (B1:◁ and B2:▷); (3) SiON (C1:△) IL formation in NCVD samples grown from $Hf(NO_3)_4$; (4) Pre-growth of a $Si_3N_4$ IL followed by $HfO_2$ growth by NCVD (C2:◇). There are several interesting aspects: (1) The $D_{it}(E)$ distributions in N-free ALCVD (D:

▽) samples are symmetric with respect to the midgap point (as also demonstrated in an enlarged view in Fig. 2) and exhibit two peaked features, one in the lower and one in the upper part of the Si band gap assigned to the (+/0) and (0/-) transitions of the $P_{b0}$ defect, respectively.[6] In MOCVD samples (data not shown here), the density of interface states is slightly higher than in ALCVD ones, but the $D_{it}$ values are the same within 15%, experimental accuracy limit. (2) As compared to the N-free entities, the growth of a thin $Si_3N_4$ (~0.3 nm) interlayer is seen to diminish $D_{it}$ below the midgap (A1:○; B1:◁) with attendant wiping out of $P_{b0}$-related peak. With respect to the 0.35 nm-$Si_3N_4$ IL, the thicker interfacial nitride (1.2 nm) in sample B2 (▷) is seen to result in a larger $D_{it}$, but the trap density still remain lower than in N-free ALCVD samples. Introducing, during CVD, nitrogen to the HTB by NO admixture (A1:○; A2:□) leads to a higher trap density in the upper part of the Si bandgap than for samples grown without NO, suggesting formation of additional acceptor-like interface traps above Si midgap. These A1 and A2 samples exhibit an asymmetric $D_{it}$ distribution. An example of a pronounced asymmetry in $D_{it}(E)$ between the lower and upper part of Si bandgap, observed in the samples A1(○) is presented in Figs. 1, 2. As compared to MOCVD samples with the thinnest $Si_3N_4$ IL (B1:◁) pre-growth of a "thicker" (1.16 nm) $Si_3N_4$ IL significantly increases $D_{it}$, particularly in the upper part of the gap. The $D_{it}$ distributions for the samples A1(○) and A2 (□) are rather smooth and reveal no peaked features. The asymmetry of the $D_{it}$ distributions in these CVD samples correlates with the addition of the NO to the hafnium precursor [cf. the samples A1:○, B1:◁, and D:▽ in Fig.2]. (3) By contrast, in the NCVD sample C1(△) in Fig. 1 the asymmetry in $D_{it}$ is much smaller than observed in the CVD samples in HTB+NO. Further, the pre-growth of thicker interfacial nitride (C2:◇) prior to the NCVD process results in the highest $D_{it}$ throughout the whole energy gap, yet with a slight asymmetry: The interface trap density in the lower part of the gap is somewhat higher than in the upper part. In this case, the asymmetry in $D_{it}(E)$ likely arises from a higher density of donor-like interface traps.

A high density of interface traps in samples with a $Si_3N_4$ IL (cf. Fig.1) could result in considerable fluctuations of the surface potential affecting the $D_{it}$ extraction. This prompts us to carry out a more detailed analysis of $G/\omega$ curve. The broadening parameter $\sigma_s$, characterizing surface potential fluctuations, is determined from the $G/\omega$ vs $log(f)$ plot as the ratio

$$\sigma_s = \frac{G_p/\omega_{f\,max/5}}{G_p/\omega_{f\,max}}, \qquad (1)$$

where $f = \frac{\omega}{2\pi}$ is the frequency, and $G_p/\omega_{f\,max}$, and $G_p/\omega_{f\,max/5}$ are real parts of interface trap admittance taken at the peak frequency $f_{max}$ and $f_{max/5}$, respectively.[11] However, typical values of $\sigma_s$ obtained for the (100)Si/SiO$_2$ interface (1.8-2 at the peak $D_{it}$ values) appear to be slightly higher than those observed here for the samples with a thick $Si_3N_4$ IL (1.25-1.5) indicating that the lateral nonuniformity of the oxide charge possibly arising from a nonuniform distribution of trapping centers is unlikely. Also, the broadening parameter decreases as $D_{it}$ increases, which could physically mean the screening of surface potential fluctuations by the charge of interface states. With increasing $D_{it}$, the ac conductance probe can sense interface states closer to the band edges because the values of the surface potential in accumulation may not saturate versus applied voltage, thus making it possible to detect an ac frequency response of interface traps [cf. the samples A1:○, A2:□, B1:◁, B2:▷ and C1:△, C2:◇ in Fig.1]. A similar observation was reported previously for the case of MOS-structures with a tunnel-thin insulator.[14]

It appears that the density of N-related interface traps in as-grown Si/IL/HfO$_2$ samples is dependent on the applied deposition process. Likely, the key factor here is the degree of Si surface oxidation, i.e. the Si/SiO$_x$ interface formation. It is not possible, though, to isolate the most influential factor(s) as indeed, growth temperature, pressure, and presence of different impurities (e.g., H) may all affect transport of oxidizing spices through the growing oxide film. However, what the data so far show is that the occurring $D_{it}$ density is dependent on (i) the thickness of the pre-grown $Si_3N_4$ IL and (ii) the method of nitrogen incorporation in the $HfO_2$ film. The latter may be inferred from comparison $D_{it}$ data for the CVD and NCVD processes. Concerning the NCVD process, deposition of $HfO_2$ from the $Hf(NO_3)_4$ precursor occurs in a H and C free environment, which is seen to result in a higher trap density than for the CVD process using the HTB precursor [cf. Fig. 1, samples A2: □ and C2: ◇]. One possible explanation may be an enhancement of silicon oxidation by the presence of hydrogen in CVD samples, which may result in a lower trap density. The application of a $Si_3N_4$ IL would thus not prevent the oxidation of the silicon surface during deposition of hafnium oxide from HTB.

The influence of thickness of the interfacial nitride on $D_{it}$ can be also inferred from CV curves as shown in Fig. 3 for two measurements temperatures, 77K and 300K. The samples grown from HTB+NO (A1: ○, A2: □) exhibit a higher density of acceptor-like interface states than those without NO admixture (B1:◁, B2:▷), as revealed by a larger positive shift of the CV-curve taken on n-type capacitors. The shift is also distinctly larger at 77 K than at 300 K, indicating trapping of electrons by defects with the energy level below the Fermi level in Si, which cannot be thermally emited.[15] At the same time, in the samples with the thicker $Si_3N_4$ IL and $HfO_2$ grown with addition of NO (A2:□), a large negative shift of the CV-curves taken on p-type capacitors is observed, pointing to the presence of additional donor-type defects. Most likely, the additional electron and hole traps are introduced during the $HfO_2$ deposition process and may be related to the particular method of the nitrogen incorporation. Apparently then, the volume of the $Si_3N_4$ involved has also a large impact on the trap density.

The electron and hole capture cross-sections ($\sigma_{e,h}$) derived from ac conductance data are shown in Fig. 4 for different samples. There are several observations: (i) For the trap energy close to Si midgap the lowest values of hole and electron capture cross sections are observed for N-free ALCVD samples (D:▽). From thereon the values $\sigma_{e,h}$ decrease monotonically, towards the gap edges as shown in Fig. 4, is the trend observed for all studied samples. The validity of the

Shockly-Read-Hall recombination-generation mechanism for charge exchange through the interface states is evident from those plots. According to Shockley-Read-Hall statistics, [16,17] interface traps can quickly respond to a change in the surface potential at the interface, and therefore, have been termed fast interface states. Another class of trapping centers with time constants considerably longer than those of the fast states are the so-called slow traps which are usually located in the interfacial region within a few nanometer of the interface plane. For the latter traps the exchange of charge with Si may occur via tunnelling (direct and/or trap assisted) and capture probability decreases with increasing distance of the traps from the interface. Possibly, for traps located in near interfacial $HfO_2$ layer the chemically grown $SiO_x$ IL might result in the lowest values of $\sigma_{e,h}$. Indeed, previous studies of the charge transport mechanisms in $SiO_x/ZrO_2$ and $SiO_x/Ta_2O_5$ gate dielectric stacks suggested the occurrence of direct tunneling of electrons across the $SiO_x$ layer and trap-assisted tunneling of electrons through traps with energy levels below the conduction band of the high permittivity dielectric layer.[18,19]

(ii) All studied N-containing entities exhibit a lower $\sigma_h$ (A1:○ and A2:□; B1:◁ and B2:▷; C1:△ and C2:◇) than in thermally grown $(100)Si/SiO_2$ (×). Apart from a small energy range near midgap a similar trend is observed for $\sigma_e$. This points to the different origin of traps distributed throughout nitride. Additionally, the hole capture cross section is found to be only weakly dependent on the technological way of incorporation of nitrogen. All values fall within a narrow range, in particular, near the top of the valence band edge. This indicates that the pre-deposition of a 0.3-0.4 nm thick silicon nitride is sufficient to form an energy barrier for holes similar to the case thick $Si_3N_4$ layers.

(iii) The $\sigma_{e,h}$ values in nitrided samples are by 2-3 orders of magnitude lower near the band edges than encountered in $(100)Si/SiO_2$ and, furthermore, get smaller as the $Si_3N_4$ IL becomes thicker. This suggests that the higher density of interface traps stems from defects located in the nitrided IL.

(iv) The nitrogen incorporation in hafnia during the CVD process using HTB+NO reduces the electron capture cross section $\sigma_e$ by one order of magnitude if the interfacial $Si_3N_4$ IL is thin [cf. samples A1:○ and B1:◁]. However, NO admixture appears to have little or no impact on $\sigma_e$ in the samples with a thick pre-grown $Si_3N_4$ IL [cf. samples A2:□ and B2:▷]. This indicates that, if the pre-grown interfacial nitride is only a few tenths of a nm thick, the method of nitrogen incorporation in subsequently deposited hafnia layer can still affect the density and spatial distribution of interface traps. However, if the pre-grown $Si_3N_4$ is rather thick (i.e. ≥1.2 nm), it dominates the interface trap density, whether or not the hafnia on top contains nitrogen.

**B. Effect of PDA**

The impact of the oxidizing PDA on $D_{it}$ is illustrated in Fig. 5, showing $D_{it}$ data inferred from ac conductance measurements (panel a) and 100 Hz CV observations after corona charging (panel b). Here, we first observe that the ALCVD (100)Si/HfO$_2$ interface (D:∇) attains the lowest $D_{it}$ due to thermal growth of an SiO$_2$ layer.[8] Next comes the MOCVD sample (E:+) with nominally the same SiO$_x$ thickness as the ALCVD one: A higher density of states is observed, but still lower than found in the N-containing samples. Among the latter most efficient reduction of the $D_{it}$ is observed in samples with the thicker $Si_3N_4$ interlayer (A2: □, B2:▷, C2:◇) as compared to the as-grown case [cf. Fig.1 and Fig.5], indicating that some oxidation of Si has occurred. This suggests that 1.2-1.5 nm thick $Si_3N_4$ IL cannot suppress the oxygen diffusion entirely. The substantially reduced density of interface traps does suggest that the thick interfacial $Si_3N_4$ IL tends to form nitrogen-containing hafnium silicate interlayer during thermal annealing as was found in the case of HfAlO$_x$/SiN/Si(100) structure after annealing in high vacuum.[20]

Below the midgap, in the energy range where the contribution of fast interface states is dominant, [cf. ac conductance data in Fig.5(a)], the $D_{it}$ values of NCVD samples (C1:△) and

CVD samples grown with and without addition of NO (A1:○ and B1:◁) are very comparable, which fact is corroborated with the $D_{it}$ results obtained using the CV measurements [Fig.5(b)]. More difference in $D_{it}$ values is observed near and above Si midgap [Fig.5(a)]. The lowest $D_{it}$ is observed for samples with an SiON interlayer (C1:△) at around 0.4-0.56 eV from the top of the valence band. By contrast, in the upper part of the gap the NCVD sample (C1:△) $D_{it}$ data fall in between those of containing nitrogen in hafnia (A1:○, A2:□, C2:◇], and without it (B1:◁, B2:▷).

Another finding is that in the all N-containing samples the $D_{it}$ values obtained from CV curves appear to be close near the band edges. A substantial impact of the processing is observed only in the midgap region, where the data exhibit a common bump-like feature (at around $E\sim E_V+0.45$ eV). The latter suggests the presence of a common defect in concentrations sensitive to the HfO$_2$ deposition conditions. The results of Schmidt *at al.* on PECVD grown Al/SiN/(100)p-Si capacitors also indicated the presence of three types of defects with a very broad density distribution at around 0.4-0.5 eV above the edge of the silicon valence band.[22] Despite of some differences between the deposition methods (PECVD and CVD), as a common feature, the nitrogen incorporation obviously produces additional type of defects near the midgap, located within thin N-containing IL. Observed results show that in the samples subjected to PDA the density of donor-like interface traps is mostly dependent on the thickness of interfacial nitride, while the density of acceptor-like interface traps is sensitive to the presence of nitrogen in hafnia film itself for all the methods of nitrogen incorporation.

The presence of additional, likely N-related, defects in the lower part of the gap may also be inferred from Fig.6, where hole and electron capture cross section data are plotted versus energy for the samples subjected to PDA. In terms of capture cross-sections, PDA of ALCVD hafnia grown on (100)n-Si yields interface cpmparable to (100)Si/SiO$_2$(x): electron capture cross section corresponding to fast interface states coincide within 50% for both samples in the energy range of 0.8-0.9 eV. The $\sigma_h$ values in N-free (100)Si/ALD HfO$_2$ entity (○) are by one order of magnitude lower than in Si/SiO$_2$(x). Furthermore, capture cross sections are found to be sensitive to nitrogen content: (i) generally, much reduced $\sigma_h$ values are observed in the CVD samples grown without addition of NO in HTB B1(◁) and B2(▷) are matching within 10%, while they are substantially differ in the samples A1(○) andA2(□) in the energy range of 0.25-0.4 eV. (ii) the electron capture cross section depends on the thickness of interfacial nitride, but not on NO admixture to the HTB: almost close $\sigma_e$ values are observed in the samples A1(○), B1(◁) and A2(□), B2(▷) in the energy range of 0.65-0.75 eV. Further, in the energy range towards the conduction band edge no correlation of $\sigma_h$ values with the method of nitrogen incorporation can be done, because admittance measurements yield two values of capture cross section.

Two groups of traps with different $\sigma_e$ are observed both in ALCVD sample D(▽) and in all the studied N-containing samples grown by CVD: A1(○),A2(□), B1(◁), B2(▷). However, for the latter, $\sigma_e$ values in the midgap region are substantially lower than those in the N-free sample D(▽). In passing, we futher note, that both electron [~(2.3-2.7)·10$^{-17}$ cm$^2$ at 0.75-0.85 eV] and hole[~(3.3-4.3)·10$^{-17}$ cm$^2$ at 0.28-0.4 eV ] capture-cross-sections found here in the ALCVD sample D(▽) are in very good agreement with those reported previously for Hf-silicate films.[23] Silicate formation upon PDA may result in a different permittivity of IL, probably consisting of (Si–O–Si) SiO$_2$ and (Si–O–Hf) hafnium silicate as found [24] for the films prepared on SiO$_2$ underlayer. Hither dielectric constant of IL may explain why the extracted values of capture cross sections are approximately 3-4 times lower than those of the dangling bonds $P_{b0}$ centers [~(10$^{-16}$ cm$^2$ at 0.8-0.85 eV] and [~(1.0-2.0)·10$^{-16}$ cm$^2$ at 0.28-0.35 eV] at the (100)Si/SiO$_2$ interface [cf. sample (×) in Fig.5]. Very close values of capture cross sections of $P_{b0}$ defect [~2·10$^{-16}$ cm$^2$ at 0.31 eV] was found earlier [26] using temperature dependent charge pumping technique in (100)Si/SiO$_2$ samples.

Two different branches in $\sigma_{e,h}(E)$ are observed after PDA in the energy range of 0.25-0.45 and 0.7-0.9 eV, which have not been encountered in as-deposited N-containing samples. For instance, in samples B1($\triangleleft$) and B2($\triangleright$) we found two distinctly different electron capture cross sections in the energy range of 0.8-0.9 eV, (cf. Fig. 6), corresponding to two peaks observed in the $G/(w\cdot A)$ vs $log(f)$ plots in accumulation shown in Fig.7. Here, $G/(w\cdot A)$ represents the real part of the ac conductance, where $A$ is the gate area. Two peaks in the $G/(w\cdot A)$ vs $log(f)$ plot are found in both p- and n-type capacitors, but they are resolved more clearly in the n-type samples. Although no such double-peak structure did appear in $G/(w\cdot A)$ vs $log(f)$ plots of N-free (100)Si/HfO$_2$ samples or in thermally grown (100)Si/SiO$_2$, a broadened high-frequency portion in the $G/(w\cdot A)$ vs $log(f)$ plot has been observed in these samples (not shown here). This refers to the co-existing of two defects in the overlapped energy range.

Interestingly, two defects with two different capture cross sections were similarly observed in the energy range of 0.6-0.7 eV above the top of silicon valence band for case of (100)Si/SiO$_2$ Al-gated MOS capacitors received post-metallization anneal in forming gas and further subjected to x-ray irradiation.[22] Two electron capture cross sections were found to depend differently on energy. One decreases almost linear towards conductance band edge, while another is weak function of energy and exists only over narrow energy range of ~100 meV. Taking into account specific of sample preparation applied there, such behaviour of electron capture cross section might be related to hydrogen induced defects.[*] In our work the samples were subjected to H-free PDA, so, the substantial amount of H-related traps in highly unlikely. Therefore, the two observed capture cross sections capture cross sections might be ascribed to the effect of nitrogen. Here, we found: (i) the peaks occure at strongly different frequencies, suggesting a large difference in cross sections [see in Fig.7]; (ii) broadening parameter $\sigma_s$ of 2.3-2.5, which correspond to surface potential fluctuations of 60-65 meV at 300 K. This does suggest double-peak structure in the $G/(w\cdot A)$ vs $log(f)$ plot to be not due to lateral charge nonuniformities, but rather refur to the presence of two types of defects with distinctly different capture cross sections. Whether or not one of these two kinds of defects could be ascribed to $P_{b0}$ centers or some other type of imperfections remains to be studied. However, it is clear that, in the sense of the capture kinetics (100)Si/HfO$_2$ interface, even being subjected to PDA is not (100)Si/SiO$_2$-like.

**3.3. Samples passivated in H$_2$**

The $D_{it}$ data obtained after hydrogen passivation at 400 $^0$C are shown in Fig.8 (a, b). In the N-containing samples, the asymmetry in the trap distributions between lower and upper halfs of silicon band gap is observed similarly to the as-grown case. It is worth to notice that in the N-free samples distribution of interface traps is almost symmetric with respect to the midgap. The minimum interface state density is found below the midgap for the samples A1(○), B1(◁) and C1 (△), while a slightly higher density of interface states is observed in the samples D (▽). The use of a thicker Si$_3$N$_4$ interlayer (B2:▷) apparently hinders passivation of fast interface states, leading to a higher $D_{it}$. Comparing data for CVD samples A1 (○) and B1 (◁) one can conclude that the addition of NO to the hafnium precursor does not impede passivation of interface states below midgap provided that the Si$_3$N$_4$ layer remains thin. Furthermore, NO seems to promote passivation of interface traps in the upper part of Si band gap. Thus, the concept of monolayer nitrogen incorporation introduced by Lucovsky [27], being applied to the high-κ case shows the beneficial aspects of nitrogen incorporation: Interface trap density as low as in N-free samples can be achieved by using pre-growth of few monolayer thick Si$_3$N$_4$ in ammonia.

Interestingly, there is a correlation between the degree of passivation of interface states in hydrogen and values of capture cross sections obtained after PDA. In the samples with close values of capture cross sections passivation results in similar values of $D_{it}$ [cf. samples A1(○) and

D (▽) in Fig.6 and in Fig 8(a) below the midgap]. Despite that density of interface states in samples A1(○) and D (▽) differ after PDA, subsequent passivation in hydrogen results in almost similar $D_{it}$ values. Similar observation is made on the other N-containing samples grown by CVD: Lower values of electron capture cross sections obtained after PDA correspond to the higher density of interface traps appeared after passivation. Apparently, nitrogen incorporation during CVD process affects interface state density in the upper part of silicon band gap. Concerning the NCVD process, highest $D_{it}$ values are seen below the midgap in the sample C2 (◇). The CVD process (sample A2: □), by contrast, results in $D_{it}$ values by almost one order of magnitude lower, although in both samples pre-grown $Si_3N_4$ IL is of the same thickness. This could be a consequence of different oxidation conditions of silicon surface during CVD and NCVD processes. In the latter case no additional $O_2$ supply is available at interface during subsequent growth of hafnia film, thus preserving Si substrate from oxidation. The presence of hydrogen in CVD process might lead to formation of oxinitride IL instead of pure $Si_3N_4$, thus enhancing hydrogen permeability during passivation.

As compared to the effect of treatment in $H_2$ at 400 $^0$C, the increase of the passivation temperature to 550 $^0$C results in a higher interface trap density in all the samples as is evident from Fig. 9(a, b). This increase in the total interface trap density (traced by both the CV and conductance method) is accompanied by an increase in a peak-like feature located at 0.4-0.5 eV above the Si valence band edge. This $D_{it}$ is quite dissimilar to the $P_{b0}$-related two-peak pattern observed in the N-free samples [cf. Fig.1 and Fig. 4]. This fact together with resistance of interface states to passivation in hydrogen suggest that the Si dangling bond defects cannot account for additional $D_{it}$ in the N-containing samples. Then it is likely, that these traps are associated with electron states of defects created by nitrogen.

## IV. Conclusions

The present work allows us to draw several major conclusions conserning influence of nitrogen on interface traps in (100)Si/HfO$_2$ structures: (i)incorporation of N to the IL or HfO$_2$ itself results in general increase of interface trap density across the ebtire Si bandgap. This Dit enhancement is due to both enchanced density of Si dangling bond defects ($P_{b0}$- centers and slow insulator related states. (ii) the presence of Si$_3$N$_4$ IL at the (100)Si/HfO$_2$ interface results in significant additional contribution to the trap density, which is still observed even after oxidazing PDA and passivation in hydrogen. (iii) In addition to continuum of interface trap levels in Si bandgap, N-related centers give a distinct peak in Dsit at 0.4-0.5 eV above the toop of Si valence band, suggesting a well defined defect structure. (iv) The N-related interface traps exhibit substantial resistance to passivation by hydrogen particularly in structures with relatively thick (~1 nm) Si$_3$N$_4$ interlayer. (v) A thin (0.3-0.4 nm) Si$_3$N$_4$ IL allows efficient passivation of interface traps at 400 $^0$C enabling one to reach $D_{it}$ values close to those observed in N-free Si/HfO$_2$ samples. Increase of the passivation temperature to 550 $^0$C leaves significant density of traps unpassivated.

Table 1. Deposition parameters and composition of (100)Si/interlayer/HfO$_2$ samples

| Sample (symbols in Fig.1-8) | Growth method | Interfacial layer thickness (nm) | CVD precursor | Oxide deposition temperature ($^0$C) | HfO$_2$ thickness (nm) |
|---|---|---|---|---|---|
| A1 (○, ●) | CVD | 0.35 Si$_3$N$_4$ | HTB*+(NO) | 400 | 7-12.5 7-12.5 |
| A2 (□, ■) | | 1.16 Si$_3$N$_4$ | HTB+(NO) | 400 | |
| B1 (◁, ◀,) | | 0.31 Si$_3$N$_4$ | HTB | 400 | 8 |
| B2 (▷, ▶) | | 1.2 Si$_3$N$_4$ | HTB | 400 | 8 |
| C1 (△, ▲) | NCVD | 1.5 SiON | Hf(NO$_3$)$_4$ | 350 | 7-12 |
| C2 (◇, ◆) | NCVD | 1.2 Si$_3$N$_4$ | Hf(NO$_3$)$_4$ | 350 | 7-12 |
| D (∇, ▼) | ALCVD | 0.7 SiO$_x$ 0.7 SiO$_x$ | HfCl$_4$ | 300 | 5 |
| E (+) | MOCVD | | Tetrakisdiethyl aminohafnium | 485 | 5 |
| (100)Si/SiO$_2$ (✕) | dry oxidation | | | 800 | 6.5 |

*hafnium tert-butoxide

Figure captions.

Figure 1. Interface state energy distributions for as-grown (100)Si/HfO$_2$ entities; samples A1:○; A2:□; B1:◁; B2:▷; C1:△; C2:◇; D:▽ are described in Table 1. $D_{it}$ values in lower and upper part of Si bandgap are determined from measurements on p- and n-type capacitors, respectively. The origin of the energy scale is placed at the Si valence band top $E_V$.

Figure 2. Example of asymmetry in interface states energy distribution brought by nitrogen in (100)Si/SiO$_2$ entities. The samples are described in caption of Fig.1.

Figure 3. Capacitance–voltage (CV) curves recorded at 100 kHz for as-grown (100)Si/HfO$_2$ entities; samples (a) B1: (◁,◀) and A1: (○,●); (b) B2: (▷,▶) and A2: (□,■) as described in Table 1. Open symbols represent CV-curves measured at 300 K, closed symbols represent CV-curves recorded at 77 K.

Figure 4. Electron and hole capture cross sections for as-grown (100)Si/HfO$_2$ entities as determined from ac conductance data; samples A1:○; A2:□; B1:◁; B2:▷; C1:△; C2:◇; D:▽ are described in Table 1. The origin of the energy scale is placed at the Si valence band top $E_V$.

Figure 5: Interface state energy distributions in (100)Si/HfO$_2$ entities subjected to PDA. (a) determined from ac conductance data; (b) determined from 100 Hz CV traces by using corona charging; samples A1:○; A2:□; B1:◁; B2:▷; C1:△; C2:◇; D:▽; E:+ are described in Table 1. The origin of the energy scale is placed at the Si valence band top $E_V$.

Figure 6. Electron and hole capture cross sections for (100)Si/HfO$_2$ entities subjected to PDA in N$_2$+O$_2$ at 800 $^0$C; samples A1:○; A2:□; B1:◁; B2:▷; C1:△; C2:◇; D:▽ and (100)/SiO$_2$ (✕) are described in Table 1. The origin of the energy scale is placed at the Si valence band top $E_V$.

Figure 7. Normalized by contact area $A$ ac conductance ($G/\omega \cdot A$) vs. frequency ($f$) curve for as-grown sample A1. Surface potential $\psi_s$ is parameters. Silicon substrate is n-type.

Figure 8: Interface state energy distributions in (100)Si/HfO$_2$ entities subjected to PDA and subsequent annealing in H$_2$ at 400 $^0$C. (a) determined from ac conductance data; (b) determined

from 100 Hz CV traces; samples A1: ○; A2:□, ■; B1: ◁; C1: △, ▲; C2: ◆; D: ▼ are described in Table 1. Open and filled symbols correspond to results obtained in p-type and n-type MOS capacitors, respectively. If necessary, corona charging of the capacitors was applied prior CV curve recording, cf. Ref. 13. The origin of the energy scale is placed at the Si valence band top $E_V$.

Figure 9: Interface state energy distributions in (100)Si/HfO$_2$ entities subjected to PDA and subsequent annealing in H$_2$ at 550 $^0$C. (a) determined from ac conductance data; (b) determined from 100 Hz CV traces; samples A1:○; A2:□; B1:◁,◀; B2:▷,▶; C1:△,▲; D:▼ are described in Table 1. Open and filled symbols correspond to results obtained in p-type and n-type MOS capacitors, respectively. If necessary, corona charging of the capacitors was applied prior CV curve recording, cf. Ref. 13.The origin of the energy scale is placed at the Si valence band top $E_V$.

Figure 1. Y. Fedorenko at al.

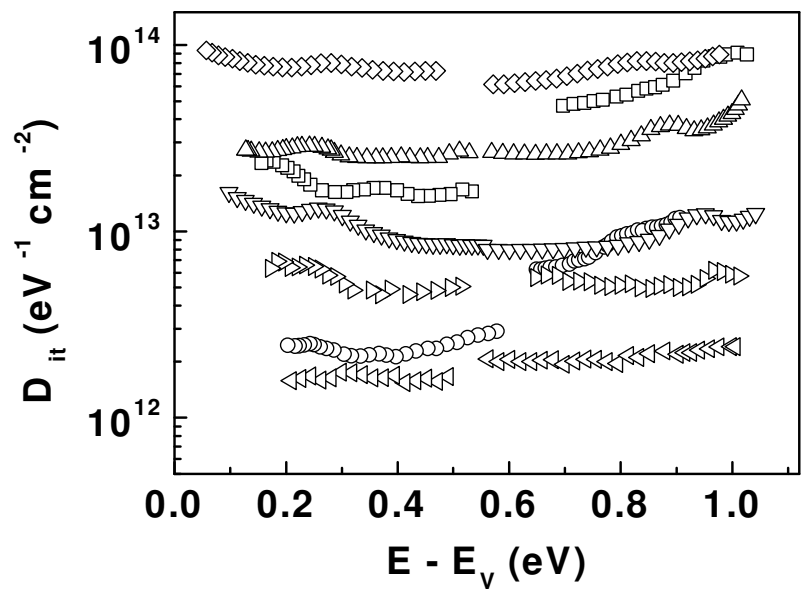

Figure 2. Y. Fedorenko at al.

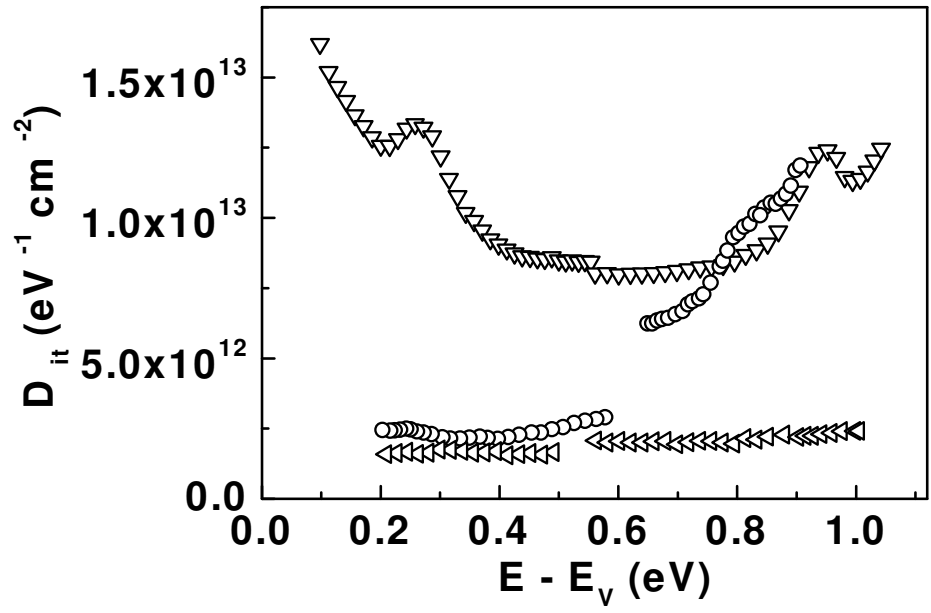

Figure 3. Y. Fedorenko at al.

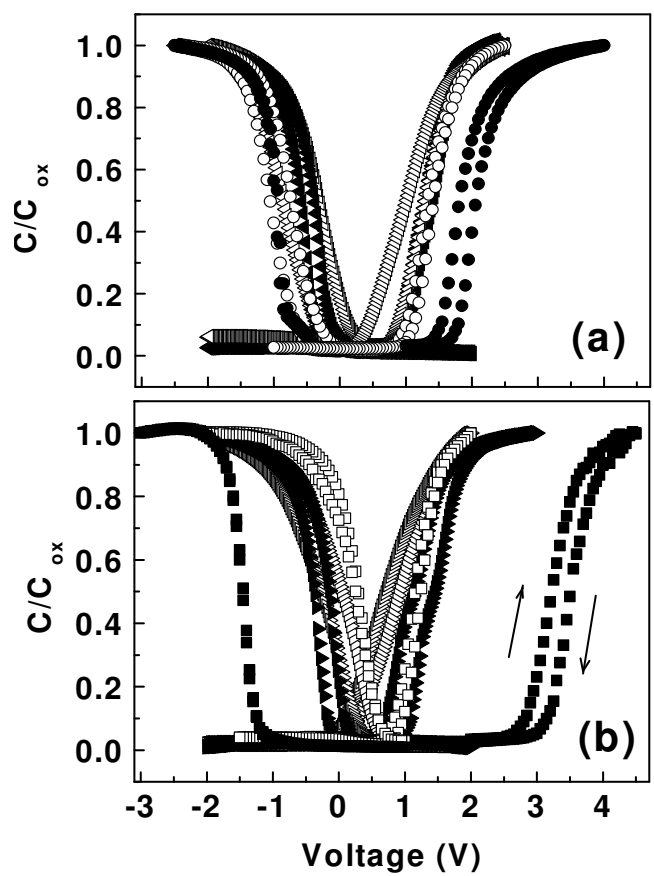

Figure 4. Y. Fedorenko at al.

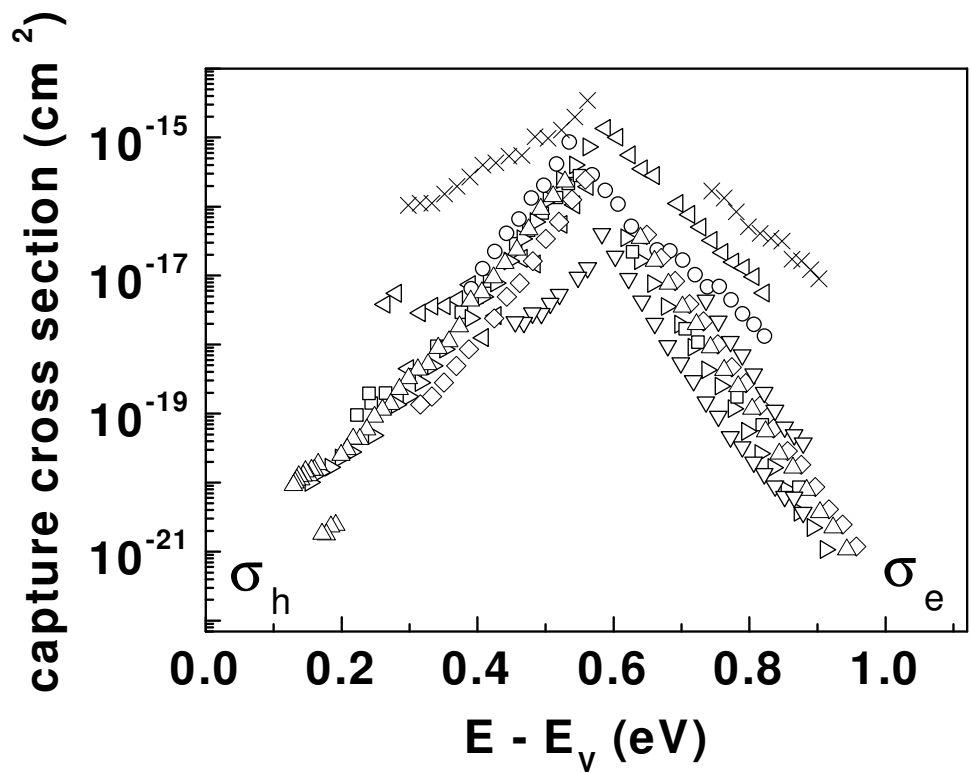

Figure 5. Y. Fedorenko at al.

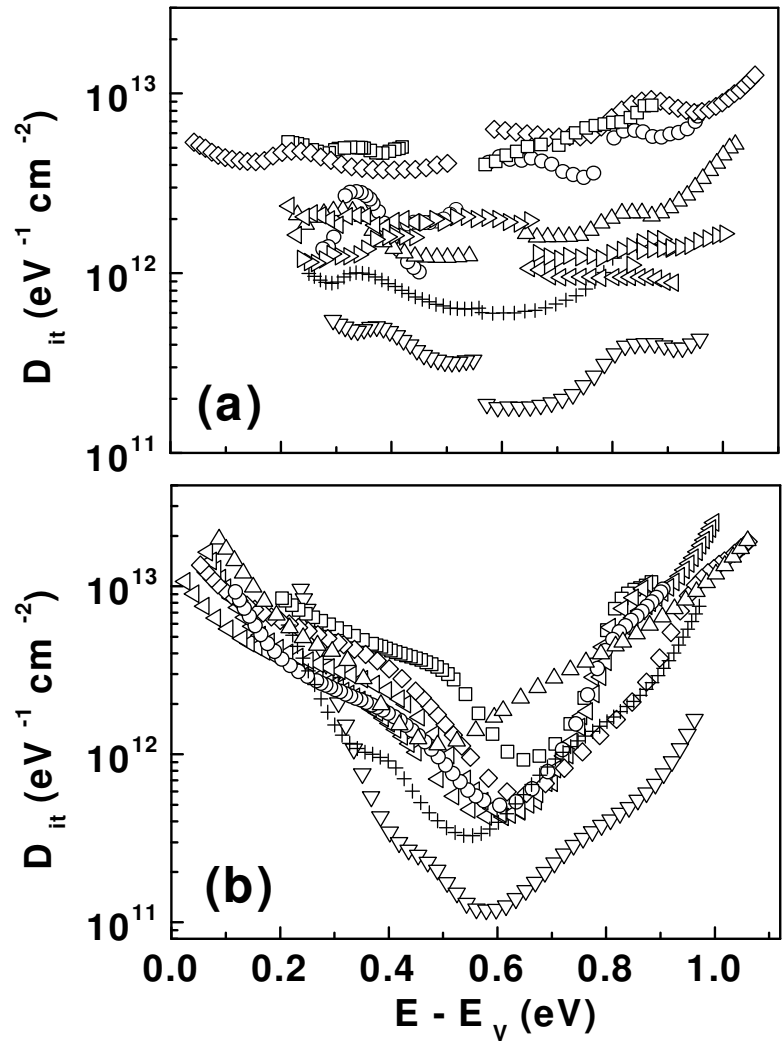

Figure 6. Y. Fedorenko at al.

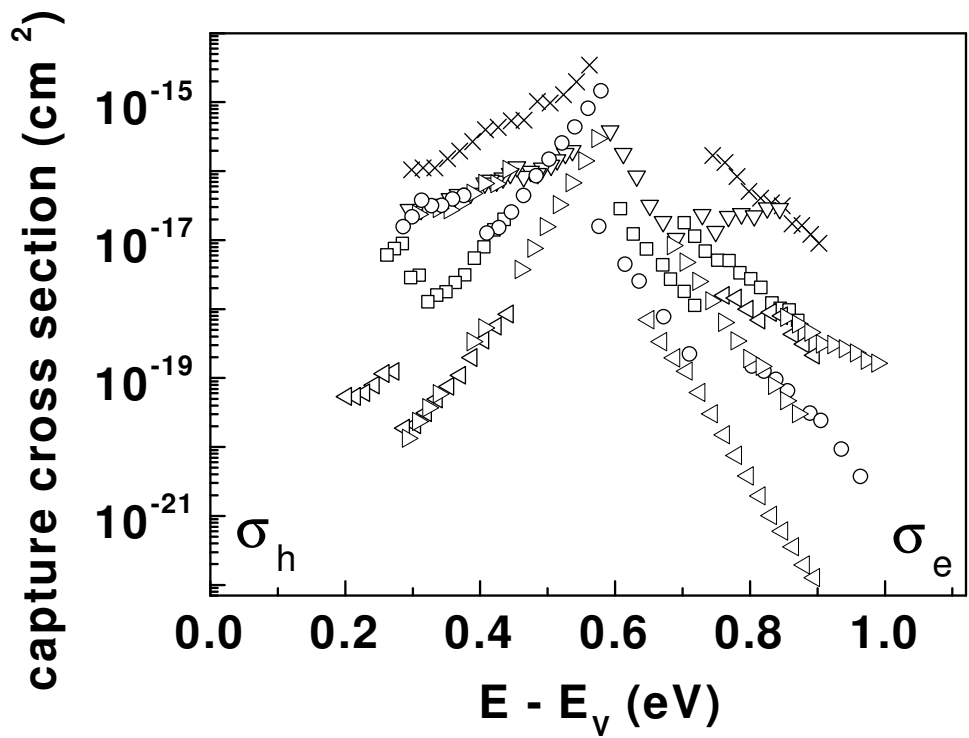

Figure 7. Y. Fedorenko at al.

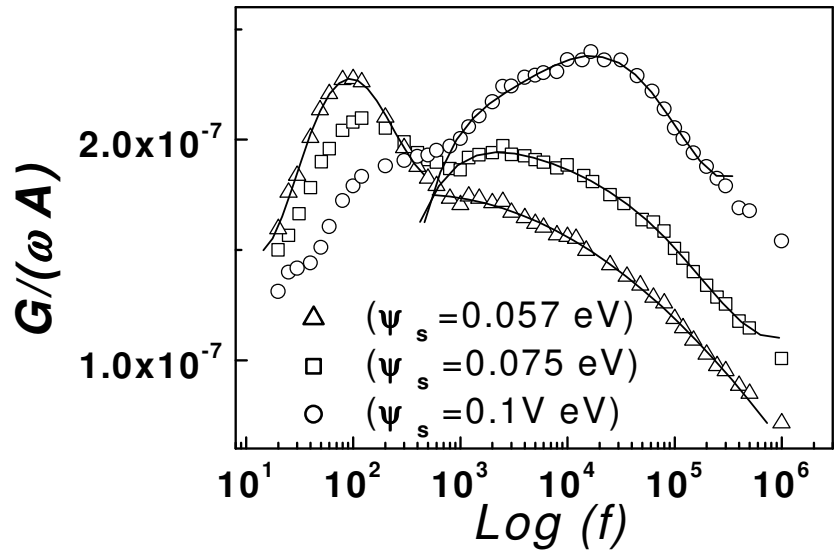

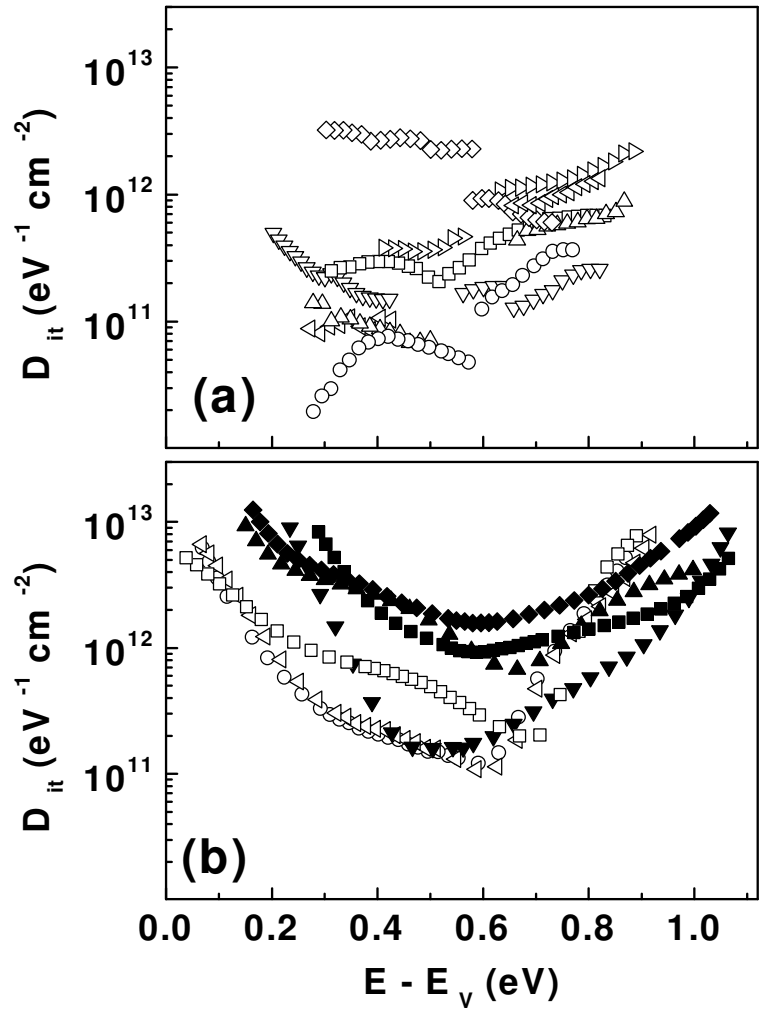

Figure 9. Y. Fedorenko at al.

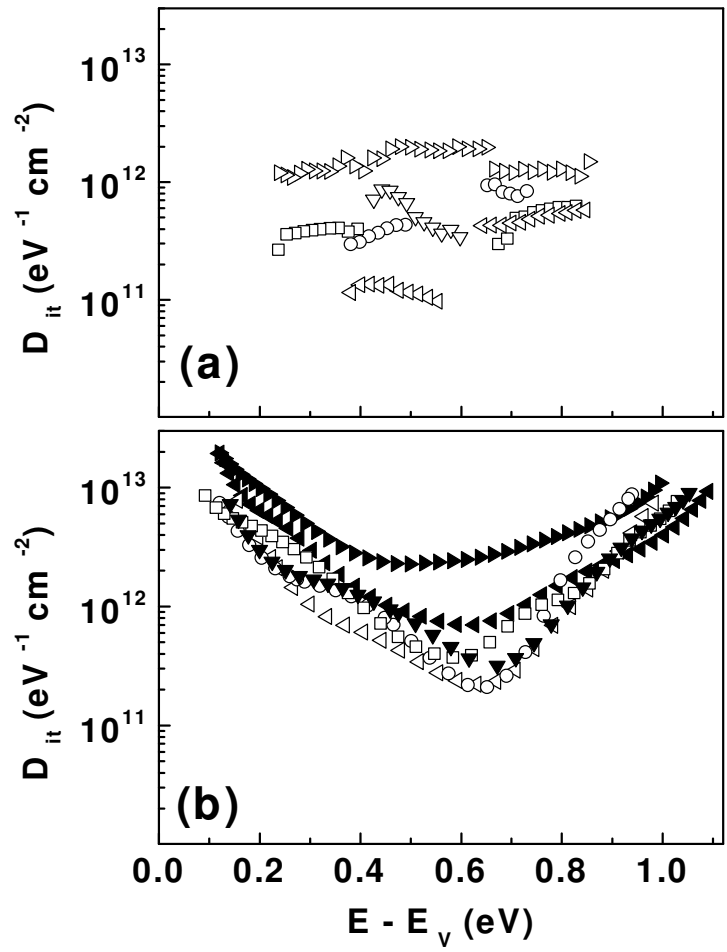